\documentclass[amsmath,showpacs,nofootinbib,11pt,superscriptaddress]{revtex4-2}
\usepackage{graphicx}
\usepackage{dcolumn}
\usepackage{bm}
\usepackage{subcaption}
\usepackage{color} 
\usepackage{slashed}
\usepackage{float}
\usepackage{amsfonts}
\usepackage{multirow}
\usepackage{lineno}
\usepackage[colorlinks=true,linkcolor=blue,citecolor=blue,urlcolor=blue]{hyperref}

\begin{document}

\newcommand{\hs}{\hspace*{0.5cm}}
\newcommand{\vs}{\vspace*{0.5cm}}
\newcommand{\be}{\begin{equation}}
\newcommand{\ee}{\end{equation}}
\newcommand{\bea}{\begin{eqnarray}}
\newcommand{\eea}{\end{eqnarray}}
\newcommand{\ben}{\begin{enumerate}}
\newcommand{\een}{\end{enumerate}}
\newcommand{\bde}{\begin{widetext}}
\newcommand{\ede}{\end{widetext}}
\newcommand{\nn}{\nonumber}
\newcommand{\crn}{\nonumber \\}
\newcommand{\Tr}{\mathrm{Tr}}
\newcommand{\non}{\nonumber}
\newcommand{\noi}{\noindent}
\newcommand{\al}{\alpha}
\newcommand{\la}{\lambda}
\newcommand{\bet}{\beta}
\newcommand{\ga}{\gamma}
\newcommand{\va}{\varphi}
\newcommand{\om}{\omega}
\newcommand{\pa}{\partial}
\newcommand{\+}{\dagger}
\newcommand{\fr}{\frac}
\newcommand{\bc}{\begin{center}}
\newcommand{\ec}{\end{center}}
\newcommand{\Ga}{\Gamma}
\newcommand{\de}{\delta}
\newcommand{\De}{\Delta}
\newcommand{\ep}{\epsilon}
\newcommand{\varep}{\varepsilon}
\newcommand{\ka}{\kappa}
\newcommand{\La}{\Lambda}
\newcommand{\si}{\sigma}
\newcommand{\Si}{\Sigma}
\newcommand{\ta}{\tau}
\newcommand{\up}{\upsilon}
\newcommand{\Up}{\Upsilon}
\newcommand{\ze}{\zeta}
\newcommand{\ps}{\psi}
\newcommand{\Ps}{\Psi}
\newcommand{\ph}{\phi}
\newcommand{\vph}{\varphi}
\newcommand{\Ph}{\Phi}
\newcommand{\Om}{\Omega}

\newcommand{\AdrHEPC}{Phenikaa Institute for Advanced Study, Phenikaa University, Duong Noi, Hanoi 100000, Vietnam}

\title{Generation-separated hypercharges and dark charges: origin of flavor, neutrino masses, and dark matter}

\author{Duong Van Loi}

\email{loi.duongvan@phenikaa-uni.edu.vn (corresponding author)}
\affiliation{\AdrHEPC}

\date{\today}

\begin{abstract}
We study a framework with generation-separated hypercharges and dark charges. After the extended hypercharge symmetry $U(1)_{Y1}\otimes U(1)_{Y2}\otimes U(1)_{Y3}$ is spontaneously broken to the Standard Model $U(1)_Y$, hierarchical masses of charged fermions and small quark mixing can arise from higher-dimensional operators involving hyperon fields. Several realizations with different hyperon sectors are presented to illustrate possible flavor structures. Focusing on the realization with four hyperons, we examine neutrino mass generation through either a seesaw or a scotoseesaw mechanism, which can accommodate tiny neutrino masses and large neutrino mixing. The breaking of the dark gauge symmetry $U(1)_D$ leaves a residual $\mathbb{Z}_2$ symmetry that stabilizes dark matter, allowing either fermionic or scalar candidates with viable TeV-scale parameter regions consistent with the observed dark matter relic abundance and current direct detection constraints.
\end{abstract}

\maketitle

\section{\label{intro}Introduction}
Despite its remarkable success in describing the fundamental particles and their interactions up to the electroweak scale, the Standard Model (SM) of particle physics leaves several fundamental questions unanswered. Among the most prominent are the origin of the charged fermion mass hierarchies, the mechanism responsible for the generation of active neutrino masses, and the nature of dark matter (DM)~\cite{ParticleDataGroup:2024cfk, Kajita:2016cak, McDonald:2016ixn, Planck:2018vyg}. These open problems provide strong motivation for exploring extensions of the SM gauge structure and particle content.

Various approaches have been proposed to address these issues. The hierarchies of fermion masses has been widely studied in frameworks involving flavor symmetries, Froggatt–Nielsen mechanisms, or extensions of the gauge sector that distinguish between fermion generations~\cite{Froggatt:1978nt, Altarelli:2010gt, Ishimori:2010au, King:2013eh}. In particular, models with generation-dependent charges or non-universal Abelian gauge symmetries have attracted considerable attention as possible origins of fermion mass hierarchies and flavor structures, since the associated gauge symmetries can strongly constrain Yukawa interactions~\cite{Langacker:2008yv, He:1991qd, Babu:1996vt, Appelquist:2002mw, Grinstein:2010ve}.

Neutrino masses and DM are also commonly investigated in extensions of the SM that introduce additional gauge symmetries and new particles. In particular, the presence of right-handed neutrinos can generate neutrino masses through several mechanisms once suitable interactions are allowed, including the type-I, type-II, and type-III seesaw mechanisms, as well as radiative scenarios~\cite{Minkowski:1977sc, Yanagida:1979as, Mohapatra:1979ia, Schechter:1980gr, Magg:1980ut, Lazarides:1980nt, Foot:1988aq, Ma:2006km}. At the same time, additional gauge symmetries in the dark sector can stabilize new particles through residual symmetries after spontaneous symmetry breaking, thereby providing viable DM candidates~\cite{Bertone:2004pz, Feng:2010gw, Arcadi:2017kky, Hambye:2008bq}.

Although these directions have been extensively studied, they are often developed in separate sectors of the theory. It is therefore interesting to explore frameworks in which a common gauge structure can simultaneously influence the flavor pattern of the SM, enable neutrino mass generation, and ensure the stability of DM candidates. In this work we consider a framework in which both hypercharges and dark charges distinguish fermion generations, an idea that has recently attracted attention in the literature~\cite{Davighi:2023evx, FernandezNavarro:2023rhv,FernandezNavarro:2024hnv,VanLoi:2025fmy}. The hypercharge sector is extended to $U(1)_{Y1}\otimes U(1)_{Y2}\otimes U(1)_{Y3}$ so that each SM fermion generation is charged under its own Abelian factor. After spontaneous symmetry breaking to the SM hypercharge $U(1)_Y$, hierarchical masses of charged fermions can arise from higher-dimensional operators involving scalar singlets, referred to as hyperons. In addition, right-handed neutrinos carry generation-dependent charges under a dark gauge symmetry $U(1)_D$, allowing small active neutrino masses to be generated once appropriate interactions are introduced, while the breaking of $U(1)_D$ leaves a residual $\mathbb{Z}_2$ symmetry that stabilizes the DM candidate.

A key feature of our framework is that the same gauge structure provides a unified origin for three seemingly unrelated aspects of physics beyond the SM. In particular, the hierarchical pattern of charged-fermion masses arises naturally from higher-dimensional operators associated with the breaking of the extended hypercharge symmetry. Small neutrino masses are generated via seesaw or scotogenic mechanisms involving fields charged under $U(1)_D$. Moreover, the spontaneous breaking of $U(1)_D$ leads to an exact residual $\mathbb{Z}_2$ symmetry that stabilizes the dark matter candidate. Therefore, flavor structure, neutrino masses, and dark matter are all governed by a common underlying gauge symmetry, rather than being introduced as separate ingredients.

The paper is organized as follows. In Sec.~\ref{model} we present the gauge structure and the pattern of symmetry breaking. In Sec.~\ref{chargedmass} we discuss effective interactions responsible for charged-fermion mass hierarchies. Neutrino mass generation is examined in Sec.~\ref{neutrino} through seesaw and scotoseesaw realizations. Possible DM candidates are discussed in Sec.~\ref{dm}, and Sec.~\ref{conc} contains our conclusions.

\section{\label{model}Gauge structure with generation-separated hypercharges and dark charges}
We consider an extension of the SM gauge symmetry with the gauge group~\cite{VanLoi:2025fmy}
\be
SU(3)_C\otimes SU(2)_L\otimes U(1)_{Y1}\otimes U(1)_{Y2}\otimes U(1)_{Y3}\otimes U(1)_D,
\label{gaugesymmetry}
\ee
where the first two factors coincide with those of the SM. The next three factors correspond to a decomposition of the SM hypercharge group, $U(1)_Y\to U(1)_{Y1}\otimes U(1)_{Y2}\otimes U(1)_{Y3}$, while the last factor represents an additional dark gauge symmetry. The three generations of SM fermions are charged under $U(1)_{Y1}$, $U(1)_{Y2}$, and $U(1)_{Y3}$, respectively, and are neutral under $U(1)_D$. In addition to the SM fermions, we introduce three right-handed neutrinos. These fields are singlets under the SM gauge symmetry but carry generation-dependent charges under $U(1)_D$. They play an important role in the generation of neutrino masses and may also provide a viable DM candidate, depending on the symmetry-breaking pattern of the dark gauge symmetry. This structure allows the hypercharge sector to distinguish fermion generations, while the dark gauge symmetry plays a role in neutrino mass generation and the stability of the dark matter candidate. The fermion content and the corresponding gauge charge assignments are summarized in Table~\ref{tab:charges}.

An important feature of this construction is that the gauge anomalies associated with the hypercharge sector cancel generation by generation. Furthermore, since all SM fermions are neutral under $U(1)_D$, the only potentially nonvanishing anomalies involving the dark gauge symmetry arise from the right-handed neutrinos. In particular, the cubic anomaly $[U(1)_D]^3$ and the mixed gravitational anomaly $[\mathrm{Gravity}]^2U(1)_D$ are proportional to $\sum_i D_{\nu_i}^3$ and $\sum_i D_{\nu_i}$, where $D_{\nu_i}$ denote the $U(1)_D$ charges of the right-handed neutrinos. For the charge assignment $(D_{\nu_R},-D_{\nu_R},0)$, both sums vanish,
\be
D_{\nu_R}^3+(-D_{\nu_R})^3+0^3=0, \qquad D_{\nu_R}+(-D_{\nu_R})+0=0,
\ee
so that the cubic and gravitational anomalies cancel automatically. Without loss of generality, the charges can be normalized to $(1,-1,0)$ corresponding to the minimal integer normalization of the $U(1)_D$ charges. This assignment also ensures that the right-handed neutrinos are distinguished by the dark symmetry. Moreover, mixed anomalies involving the SM gauge groups, such as $SU(3)_C^2U(1)_D$, $SU(2)_L^2U(1)_D$, and $U(1)_Y^2U(1)_D$ vanish because all SM fermions carry zero $U(1)_D$ charge. Therefore, the gauge symmetry in Eq.~(\ref{gaugesymmetry}) is anomaly free in the fermion sector.

\begin{table}[h]
\centering
\caption{Fermion content and gauge charge assignments in the model.}
\label{tab:charges}
\begin{tabular}{c|c c c c c c}
\hline\hline
Field & $SU(3)_C$ & $SU(2)_L$ & $U(1)_{Y1}$ & $U(1)_{Y2}$ & $U(1)_{Y3}$ & $U(1)_D$ \\
\hline
$q_{1L}$ & $\mathbf{3}$ & $\mathbf{2}$ & $1/6$ & 0 & 0 & 0 \\
$u_{1R}$ & $\mathbf{3}$ & $\mathbf{1}$ & $2/3$ & 0 & 0 & 0 \\
$d_{1R}$ & $\mathbf{3}$ & $\mathbf{1}$ & $-1/3$ & 0 & 0 & 0 \\
$\ell_{1L}$ & $\mathbf{1}$ & $\mathbf{2}$ & $-1/2$ & 0 & 0 & 0 \\
$e_{1R}$ & $\mathbf{1}$ & $\mathbf{1}$ & $-1$ & 0 & 0 & 0 \\
$q_{2L}$ & $\mathbf{3}$ & $\mathbf{2}$ & 0 & $1/6$ & 0 & 0 \\
$u_{2R}$ & $\mathbf{3}$ & $\mathbf{1}$ & 0 & $2/3$ & 0 & 0 \\
$d_{2R}$ & $\mathbf{3}$ & $\mathbf{1}$ & 0 & $-1/3$ & 0 & 0 \\
$\ell_{2L}$ & $\mathbf{1}$ & $\mathbf{2}$ & 0 & $-1/2$ & 0 & 0 \\
$e_{2R}$ & $\mathbf{1}$ & $\mathbf{1}$ & 0 & $-1$ & 0 & 0 \\
$q_{3L}$ & $\mathbf{3}$ & $\mathbf{2}$ & 0 & 0 & $1/6$ & 0 \\
$u_{3R}$ & $\mathbf{3}$ & $\mathbf{1}$ & 0 & 0 & $2/3$ & 0 \\
$d_{3R}$ & $\mathbf{3}$ & $\mathbf{1}$ & 0 & 0 & $-1/3$ & 0 \\
$\ell_{3L}$ & $\mathbf{1}$ & $\mathbf{2}$ & 0 & 0 & $-1/2$ & 0 \\
$e_{3R}$ & $\mathbf{1}$ & $\mathbf{1}$ & 0 & 0 & $-1$ & 0 \\
$\nu_{1R}$ & $\mathbf{1}$ & $\mathbf{1}$ & 0 & 0 & 0 & $1$ \\
$\nu_{2R}$ & $\mathbf{1}$ & $\mathbf{1}$ & 0 & 0 & 0 & $-1$ \\
$\nu_{3R}$ & $\mathbf{1}$ & $\mathbf{1}$ & 0 & 0 & 0 & $0$ \\
\hline\hline
\end{tabular}
\end{table}

At energies below the flavor-breaking scale, scalar singlets charged under the hypercharge factors may acquire vacuum expectation values (VEVs) and spontaneously break
\be
U(1)_{Y1}\otimes U(1)_{Y2}\otimes U(1)_{Y3}
\longrightarrow U(1)_Y,
\ee
where $U(1)_Y$ is identified with the SM hypercharge symmetry. The SM Higgs doublet $H$ is assumed to carry hypercharge only under $U(1)_{Y3}$, \be H\sim(1,2,0,0,1/2,0).\ee
As a consequence, renormalizable Yukawa interactions are allowed only for the third generation of fermions, while the masses of the first and second generations arise from higher-dimensional operators involving scalar singlets that break the extended hypercharge symmetry. This structure provides a natural explanation for why the third-generation fermions are significantly heavier than those of the first two generations. 

We also introduce a scalar singlet 
\be
\chi \sim (1,1,0,0,0,2),
\ee
which carries an even $U(1)_D$ charge, so that the spontaneous breaking of $U(1)_D$ leaves an exact residual $\mathbb{Z}_2$ symmetry,
\be
U(1)_D \longrightarrow \mathbb{Z}_2.
\ee
The resulting residual $\mathbb{Z}_2$ symmetry stabilizes the DM candidate. This breaking also allows Majorana mass terms for the right-handed neutrinos $\nu_{1R}$ and $\nu_{2R}$, whose $U(1)_D$ charges are $\pm 1$. The relevant interactions are
\be\mathcal{L} \supset 
\frac{1}{2} f_1 \bar{\nu}_{1R}^c\nu_{1R}\chi^* +\frac{1}{2} f_2 \bar{\nu}_{2R}^c\nu_{2R}\chi  - m_{12}\bar{\nu}_{1R}^c\nu_{2R}+ \mathrm{H.c.},\ee
which generate Majorana masses after $\chi$ develops a VEV. 

\section{\label{chargedmass} Effective operators for charged fermion masses}
The generation-separated hypercharge structure introduced in the previous section strongly constrains the Yukawa sector. Since different fermion generations are charged under different $U(1)_{Yi}$ gauge factors, only the third-generation Yukawa interactions remain gauge invariant at the renormalizable level, as the SM Higgs doublet $H$ carries hypercharge solely under $U(1)_{Y3}$. Consequently, the masses of the first and second generations must arise from higher-dimensional operators involving scalar singlets that break the extended hypercharge symmetry. In the following, we illustrate how such operators can be generated through the exchange of scalar singlets (hyperons).

To parametrize these effects in a model-independent way, we introduce spurion fields $\Phi_i$ that compensate the hypercharge mismatch in the Yukawa interactions. Following the spurion analysis of
Ref.~\cite{FernandezNavarro:2023rhv}, the effective Yukawa Lagrangian for
charged fermions can then be written as
{\small\bea \mathcal{L} &\supset& \begin{pmatrix}
\bar{q}_{1L}\\ \bar{q}_{2L}\\ \bar{q}_{3L}
\end{pmatrix}^T\begin{pmatrix}
h^u_{11}\Ph(-1/2,0,1/2) & h^u_{12}\Ph(1/6,-2/3,1/2) & h^u_{13}\Ph(1/6,0,-1/6)\\
h^u_{21}\Ph(-2/3,1/6,1/2) & h^u_{22}\Ph(0,-1/2,1/2) & h^u_{23}\Ph(0,1/6,-1/6)\\
h^u_{31}\Ph(-2/3,0,2/3) & h^u_{32}\Ph(0,-2/3,2/3) & h^u_{33}
\end{pmatrix} \begin{pmatrix}
u_{1R} \\ u_{2R} \\ u_{3R}
\end{pmatrix}\tilde{H}\crn
&&+\begin{pmatrix}
\bar{q}_{1L}\\\bar{q}_{2L}\\ \bar{q}_{3L}
\end{pmatrix}^T\begin{pmatrix}
h^d_{11}\Ph(1/2,0,-1/2) & h^d_{12}\Ph(1/6,1/3,-1/2) & h^d_{13}\Ph(1/6,0,-1/6)\\
h^d_{21}\Ph(1/3,1/6,-1/2) & h^d_{22}\Ph(0,1/2,-1/2) & h^d_{23}\Ph(0,1/6,-1/6)\\
h^d_{31}\Ph(1/3,0,-1/3) & h^d_{32}\Ph(0,1/3,-1/3) & h^d_{33}
\end{pmatrix} \begin{pmatrix}
d_{1R} \\ d_{2R} \\ d_{3R}
\end{pmatrix}H \crn
&&+\begin{pmatrix}
\bar{\ell}_{1L} \\ \bar{\ell}_{2L}\\ \bar{\ell}_{3L}
\end{pmatrix}^T\begin{pmatrix}
h^e_{11}\Ph(1/2,0,-1/2) & h^e_{12}\Ph(-1/2,1,-1/2) & h^e_{13}\Ph(-1/2,0,1/2)\\
h^e_{21}\Ph(1,-1/2,-1/2) & h^e_{22}\Ph(0,1/2,-1/2) & h^e_{23}\Ph(0,-1/2,1/2)\\
h^e_{31}\Ph(1,0,-1) & h^e_{32}\Ph(0,1,-1) & h^e_{33}
\end{pmatrix} \begin{pmatrix}
e_{1R} \\ e_{2R} \\ e_{3R}
\end{pmatrix}H \crn
&&+ \mathrm{H.c.},\eea}
where $H$ denotes the SM Higgs doublet. The spurion structure presented above can be realized through different sets of hyperons that break the extended hypercharge symmetry. In the following, we consider three illustrative realizations with increasing numbers of hyperons, which demonstrate how the observed hierarchies of charged-fermion masses can emerge from different symmetry-breaking patterns.

The hierarchical Yukawa textures obtained in this framework admit a simple qualitative interpretation. Since different fermion generations require different numbers of hyperon insertions to restore gauge invariance, lighter generations arise from higher-dimensional operators and are therefore increasingly suppressed. This mechanism naturally leads to the observed pattern $m_1 \ll m_2 \ll m_3$ in all charged fermion sectors. This qualitative structure will be made explicit below in terms of powers of the Wolfenstein parameter $\lambda$.

Moreover, the off-diagonal entries of the Yukawa matrices typically involve more insertions than the leading diagonal ones, implying an approximate alignment of the corresponding mass matrices in flavor space. As a result, the CKM mixing angles exhibit a hierarchical structure, consistent with the observed pattern $|V_{us}| \gg |V_{cb}| \gg |V_{ub}|$. We emphasize that our goal here is to illustrate the qualitative origin of flavor hierarchies, while a detailed numerical fit is beyond the scope of this work.

\subsection{Model I with two hyperons}
As a minimal realization, we first consider the two-hyperon scenario introduced
in Ref.~\cite{FernandezNavarro:2024hnv},
\be \ph_{12}\sim(1,1,1/6,-1/6,0,0),\hs \ph_{23}\sim(1,1,0,1/6,-1/6,0). \ee
These fields connect the first and second generations, and the second and third generations, respectively. For simplicity, we assume that the cutoff scales associated with these operators are of the same order and can be characterized by a common scale $\La$. Under this assumption, the Yukawa interactions of charged fermions can be written as
\bea \mathcal{L} &\supset& \begin{pmatrix}
\bar{q}_{1L}\\ \bar{q}_{2L}\\ \bar{q}_{3L}
\end{pmatrix}^T\begin{pmatrix}
\fr{h^u_{11}}{\La^6}\ph_{12}^{*3}\ph_{23}^{*3} & \fr{h^u_{12}}{\La^4}\ph_{12}\ph_{23}^{*3} & \fr{h^u_{13}}{\La^2}\ph_{12}\ph_{23}\\
\fr{h^u_{21}}{\La^7}\ph_{12}^{*4}\ph_{23}^{*3} & \fr{h^u_{22}}{\La^3}\ph_{23}^{*3} & \fr{h^u_{23}}{\La}\ph_{23}\\
\fr{h^u_{31}}{\La^8}\ph_{12}^{*4}\ph_{23}^{*4} & \fr{h^u_{32}}{\La^4}\ph_{23}^{*4} & h^u_{33}
\end{pmatrix} \begin{pmatrix}
u_{1R} \\ u_{2R} \\ u_{3R}
\end{pmatrix}\tilde{H}\crn
&&+\begin{pmatrix}
\bar{q}_{1L}\\\bar{q}_{2L}\\ \bar{q}_{3L}
\end{pmatrix}^T\begin{pmatrix}
\fr{h^d_{11}}{\La^6}\ph_{12}^3\ph_{23}^3 & \fr{h^d_{12}}{\La^4}\ph_{12}\ph_{23}^3 & \fr{h^d_{13}}{\La^2}\ph_{12}\ph_{23}\\
\fr{h^d_{21}}{\La^5}\ph_{12}^2\ph_{23}^3 & \fr{h^d_{22}}{\La^3}\ph_{23}^3 & \fr{h^d_{23}}{\La}\ph_{23}\\
\fr{h^d_{31}}{\La^4}\ph_{12}^2\ph_{23}^2 & \fr{h^d_{32}}{\La^2}\ph_{23}^2 & h^d_{33}
\end{pmatrix} \begin{pmatrix}
d_{1R} \\ d_{2R} \\ d_{3R}
\end{pmatrix}H \crn
&&+\begin{pmatrix}
\bar{\ell}_{1L} \\ \bar{\ell}_{2L}\\ \bar{\ell}_{3L}
\end{pmatrix}^T\begin{pmatrix}
\fr{h^e_{11}}{\La^6}\ph_{12}^3\ph_{23}^3 & \fr{h^e_{12}}{\La^6}\ph_{12}^{*3}\ph_{23}^3 & \fr{h^e_{13}}{\La^6}\ph_{12}^{*3}\ph_{23}^{*3}\\
\fr{h^e_{21}}{\La^9}\ph_{12}^6\ph_{23}^3 & \fr{h^e_{22}}{\La^3}\ph_{23}^3 & \fr{h^e_{23}}{\La^3}\ph_{23}^{*3}\\
\fr{h^e_{31}}{\La^{12}}\ph_{12}^6\ph_{23}^6 & \fr{h^e_{32}}{\La^6}\ph_{23}^6 & h^e_{33}
\end{pmatrix} \begin{pmatrix}
e_{1R} \\ e_{2R} \\ e_{3R}
\end{pmatrix}H + \mathrm{H.c.},\eea 
where $v\simeq 246$~GeV denotes the VEV of the SM Higgs field.

After symmetry breaking, the hyperons develop VEVs. For illustration, we take
\be \frac{\langle\phi_{12}\rangle}{\La}\equiv\frac{v_{12}}{\La}\simeq \la,\hs \frac{\langle\phi_{23}\rangle}{\La}\equiv\frac{v_{23}}{\La}\simeq \la^{1.5},\ee
with $\la\simeq 0.224$ being the Wolfenstein parameter. The resulting fermion mass matrices take the schematic form
\be
M_u \sim v\begin{pmatrix}
\la^{7.5} & \la^{5.5} & \la^{2.5}\\
\la^{8.5} & \la^{4.5} & \la^{1.5}\\
\la^{10} & \la^{6} & 1
\end{pmatrix},\hs
M_d \sim v \begin{pmatrix}
\la^{7.5} & \la^{5.5} & \la^{2.5}\\
\la^{6.5} & \la^{4.5} & \la^{1.5}\\
\la^{5} & \la^{3} & 1
\end{pmatrix},\hs
M_e \sim v\begin{pmatrix}
\la^{7.5} & \la^{7.5} & \la^{7.5}\\
\la^{10.5} & \la^{4.5} & \la^{4.5}\\
\la^{15} & \la^{9} & 1
\end{pmatrix}.
\ee
These textures originate from the suppression of higher-dimensional operators and qualitatively reproduce the observed fermion mass hierarchies and the CKM structure $|V_{us}|\sim\la$, $|V_{cb}|\sim\la^2$, $|V_{ub}|\sim\la^3$~\cite{ParticleDataGroup:2024cfk}.

\subsection{Model II with three hyperons}
We next consider a setup with three hyperons, as introduced in
Ref.~\cite{FernandezNavarro:2023rhv}, leading to a richer structure of
effective operators and modified flavor hierarchies. The scalar fields are
given by
\be
\ph_{12}\sim(1,1,1/6,-1/6,0,0),\hs
\ph_{23}\sim(1,1,0,1/6,-1/6,0),\hs
\varphi_{23}\sim(1,1,0,1/2,-1/2,0).
\ee
These fields mediate flavor transitions among different generations and allow for a wider set of operator structures compared to Model I.

For illustration, the VEVs are now assumed as
\be
\frac{v_{12}}{\La}\simeq \la^{1.5},\hs
\frac{v_{23}}{\La}\simeq \la^{2.5},\hs
\frac{\langle\varphi_{23}\rangle}{\La}\equiv\frac{v'_{23}}{\La}\simeq \la^{3}.
\ee
The resulting fermion mass matrices can then be written schematically as
\be
M_u \sim v\begin{pmatrix}
\la^{7.5} & \la^{4.5} & \la^{4}\\
\la^{9} & \la^{3} & \la^{2.5}\\
\la^{11.5} & \la^{5.5} & 1
\end{pmatrix},\hs
M_d \sim v\begin{pmatrix}
\la^{7.5} & \la^{4.5} & \la^{4}\\
\la^{6} & \la^{3} & \la^{2.5}\\
\la^{8} & \la^{5} & 1
\end{pmatrix},\hs
M_e \sim v\begin{pmatrix}
\la^{7.5} & \la^{7.5} & \la^{7.5}\\
\la^{12} & \la^{3} & \la^{3}\\
\la^{15} & \la^{6} & 1
\end{pmatrix}.
\ee

Compared with Model~I, the additional hyperon reduces the dimensionality of several operators, leading to milder suppressions in selected entries. This yields a more flexible flavor structure, while still accommodating realistic fermion mass hierarchies and mixing patterns.

\subsection{Model III with four hyperons}
Finally, we consider the four-hyperon scenario studied in
Ref.~\cite{FernandezNavarro:2024hnv},
\begin{align}
\ph_{12}&\sim(1,1,1/6,-1/6,0,0),\hs
\ph_{23}\sim(1,1,0,1/6,-1/6,0),\\
\varphi_{12}&\sim(1,1,1/2,-1/2,0,0),\hs
\varphi_{23}\sim(1,1,0,1/2,-1/2,0),
\end{align}
which allow for a richer set of effective operators and further modify the flavor hierarchies.
Their VEVs are taken as
\be
\frac{v_{12}}{\La}\simeq\la^{2},\hs
\frac{\langle\varphi_{12}\rangle}{\La}\equiv\frac{v'_{12}}{\La}\simeq\la^{2.5},\hs
\frac{v_{23}}{\La}\simeq\la^{4},\hs
\frac{\langle\varphi_{23}\rangle}{\La}\equiv\frac{v'_{23}}{\La}\simeq\la^{4.5}.\label{eqhyper}
\ee
Then, the resulting fermion mass matrices take the schematic form
\be
M_u \sim v\begin{pmatrix}
\la^{7} & \la^{6.5} & \la^{6}\\
\la^{9} & \la^{4.5} & \la^{4}\\
\la^{13} & \la^{8.5} & 1
\end{pmatrix},\hs
M_d \sim v\begin{pmatrix}
\la^{7} & \la^{6.5} & \la^{6}\\
\la^{8.5} & \la^{4.5} & \la^{4}\\
\la^{12} & \la^{8} & 1
\end{pmatrix},\hs
M_e \sim v\begin{pmatrix}
\la^{7} & \la^{7} & \la^{7}\\
\la^{9.5} & \la^{4.5} & \la^{4.5}\\
\la^{14} & \la^{9} & 1
\end{pmatrix}.
\ee

Compared with the previous models, the presence of four hyperons further lowers the dimensionality of effective operators, leading to milder suppressions and a more flexible flavor structure. In the following, we focus on this setup and explore its implications for neutrino masses and DM.

\section{\label{neutrino}Neutrino masses}
In this section we explore how small neutrino masses can arise in Model III. We consider two representative possibilities: a conventional seesaw mechanism and a scotoseesaw scenario in which radiative corrections generate additional contributions to neutrino masses.
\subsection{Seesaw mechanism in Model III}
To realize the seesaw mechanism, we consider the construction studied in
Ref.~\cite{FernandezNavarro:2024hnv}, which involves three vector-like
messenger neutrinos
\be\nu_{12}\sim (1,1,-1/2,1/2,0,0),\hs \nu_{13}\sim (1,1,-1/2,0,1/2,0),\hs \nu_{23}\sim (1,1,0,-1/2,1/2,0), \ee
and two right-handed neutrinos $\nu_{3R}$ and
\be\nu'_{3R}\sim (1,1,0,0,0). \ee
These messenger states mediate the interactions between the SM lepton doublets and the heavy Majorana sector, thereby enabling the generation of light neutrino masses through the seesaw mechanism.

The Yukawa interactions relevant for neutrino mass generation are
\bea \mathcal{L} &\supset& h_1\bar{l}_{1L}\tilde{H}\nu_{13R} + h_2\bar{l}_{2L}\tilde{H}\nu_{23R} + h_3\bar{l}_{3L}\tilde{H}\nu_{3R} + h_4\bar{l}_{3L}\tilde{H}\nu'_{3R} + h_5\bar{\nu}_{13L}\varphi^*_{12}\nu_{23R}\crn
&&+ h_6\bar{\nu}_{23L}\varphi_{12}\nu_{13R} + h_7\bar{\nu}_{12L}\varphi_{23}\nu_{13R} + h_8\bar{\nu}_{13L}\varphi^*_{23}\nu_{12R} + h_9\bar{\nu}_{12L}\varphi_{12}\nu_{3R}\crn
&&+ h_{10}\bar{\nu}_{12L}\varphi_{12}\nu'_{3R} + h_{11}\bar{\nu}_{23L}\varphi_{23}\nu_{3R} + h_{12}\bar{\nu}_{23L}\varphi_{23}\nu'_{3R} + h_{13}\bar{\nu}_{3R}^c\varphi_{12}\nu_{12R}\crn
&& + h_{14}\bar{\nu}'^c_{3R}\varphi_{12}\nu_{12R} + h_{15}\bar{\nu}_{3R}^c\varphi_{23}\nu_{23R} + h_{16}\bar{\nu}'^c_{3R}\varphi_{23}\nu_{23R} - m_{12}\bar{\nu}_{12L}\nu_{12R} - m_{23}\bar{\nu}_{23L}\nu_{23R}\crn
&& - m_{13}\bar{\nu}_{13L}\nu_{13R} - \fr 1 2 M\bar{\nu}_{3R}^c\nu_{3R}- \fr 1 2 M'\bar{\nu}'^c_{3R}\nu'_{3R}- \fr 1 2 M_{33}\bar{\nu}'^c_{3R}\nu_{3R} + \mathrm{H.c.}, \eea
where the coefficients $h_i$ are dimensionless Yukawa couplings, while the parameters $m_{ij}$ and $M$, $M'$, $M_{33}$ have mass dimension.

After the scalar fields develop VEVs, these interactions generate Dirac and Majorana mass terms for the neutrino sector. The full neutrino mass matrix can then be written in the basis $(\nu_{aL}, \nu_{12L}, \nu_{13L}, \nu_{23L}$, $\nu_{12R}^c, \nu_{13R}^c, \nu_{23R}^c, \nu_{3R}^c, \nu_{3R}'^c)$ (with $a=1,2,3$ denoting the SM neutrinos) as
\be \begin{pmatrix}
0 & M_D \\ M_D^T & M_M
\end{pmatrix}. \ee
Following the seesaw construction of Ref.~\cite{FernandezNavarro:2024hnv}, the resulting submatrices are
\bea M_D &=& -\frac{v}{\sqrt2} \begin{pmatrix}
0 & 0 & 0 & 0 & h_1 & 0 & 0 & 0\\
0 & 0 & 0 & 0 & 0 & h_2 & 0 & 0\\
0 & 0 & 0 & 0 & 0 & 0 & h_3 & h_4 \end{pmatrix},\\
M_M &=& -\frac{1}{\sqrt2} \begin{pmatrix}
0 & 0 & 0 & m_{12} & h_7v_{23} & 0 & h_9v_{12} & h_{10}v_{12}\\ 
0 & 0 & 0 & h_8v_{23} & m_{13} & h_5v_{12} & 0 & 0\\ 
0 & 0 & 0 & 0 & h_6v_{12} & m_{23} & h_{11}v_{23} & h_{12}v_{23}\\ 
m_{12} & h_8v_{23} & 0 & 0 & 0 & 0 & h_{13}v_{12} & h_{14}v_{12} \\
h_7v_{23} & m_{13} & h_6v_{12} &0 & 0 & 0 & 0 & 0\\
0 & h_5v_{12} & m_{23} & 0 & 0 & 0 & h_{15}v_{23} & h_{16}v_{23}\\
h_9v_{12} & 0 & h_{11}v_{23} & h_{13}v_{12} & 0 & h_{15}v_{23} & \sqrt2 M & M_{33}\\
h_{10}v_{12} & 0 & h_{12}v_{23} & h_{14}v_{12} & 0 & h_{16}v_{23} & M_{33} & \sqrt2 M'
\end{pmatrix}. \eea
The Dirac mass matrix $M_D$ exhibits a sparse structure dictated by the underlying flavor symmetry, while $M_M$ contains the heavy Majorana mass terms and mixings among vector-like states. This structure realizes a multi-state seesaw mechanism, leading to suppressed light-neutrino masses.

For a numerical illustration, we take  $m_{12,13}\simeq v_{12}\simeq\mathcal{O}(10^{5})$~GeV, $m_{23}\simeq v_{23}\simeq\mathcal{O}(10^{4})$~GeV, $M\simeq M'\simeq M_{33}\simeq\mathcal{O}(10^{14})$~GeV, while all Yukawa couplings are assumed to be $\mathcal{O}(1)$. We note that these benchmark values are consistent with the hierarchical parametrization in Eq.~(\ref{eqhyper}). In particular, taking $\La\sim \mathcal{O}(10^{7})\,$GeV and $\la\simeq 0.224$, one finds $v_{12} \sim \la^{2.5}\La$ and $v_{23} \sim \la^{4.5}\La$, which correspond to $v_{12}\sim \mathcal{O}(10^{5})\,$GeV and $v_{23}\sim \mathcal{O}(10^{4})\,$GeV. Therefore, the benchmark values adopted here can be viewed as illustrative choices consistent with the underlying Wolfenstein expansion. The resulting light neutrino mass matrix is then approximately given by
\be M_\nu\simeq -M_DM_M^{-1}M_D^T\simeq \begin{pmatrix}
0.0103473 & -0.0123593 & -0.0640642\\
-0.0123593 & 0.0147853 & 0.0762653\\
-0.0640642 & 0.0762653 & 0.399516
\end{pmatrix}\text{ eV}, \ee
which has rank two and therefore predicts two nonzero neutrino masses. This feature is consistent with current neutrino oscillation data, which require at least two massive neutrinos~\cite{ParticleDataGroup:2024cfk,Kajita:2016cak, McDonald:2016ixn}. The rank-2 structure originates from the specific pattern of couplings encoded in $M_D$ and $M_M$. Although several heavy states are present, the SM neutrinos couple only to a restricted subset of them, as seen from the sparse structure of $M_D$. After integrating out the heavy sector, only two independent combinations effectively contribute to the seesaw formula, leading to a light neutrino mass matrix of rank 2 and hence one massless neutrino at leading order. In addition, different rows of $M_D$ couple to different heavy states, and these states are mixed through $M_M$. As a result, the effective mass matrix $M_\nu$ contains sizable off-diagonal entries, naturally accommodating large leptonic mixing. 

\subsection{Scotoseesaw realization of Model III}
We now consider a scenario in which the tree-level neutrino mass matrix has rank one, implying that only one neutrino acquires mass at tree level, while a second neutrino mass is generated at the loop level. In this setup, only two vector-like neutrinos, $\nu_{13}$ and $\nu_{23}$, are introduced. The Yukawa interactions relevant for neutrino mass generation are
\bea \mathcal{L} &\supset& h_1\bar{l}_{1L}\tilde{H}\nu_{13R} + h_2\bar{l}_{2L}\tilde{H}\nu_{23R} + h_3\bar{l}_{3L}\tilde{H}\nu_{3R} + h_4\bar{\nu}_{13L}\varphi^*_{12}\nu_{23R}+ h_5\bar{\nu}_{23L}\varphi_{12}\nu_{13R}\crn
&& + h_6\bar{\nu}_{23L}\varphi_{23}\nu_{3R}+ h_7\bar{\nu}_{3R}^c\varphi_{23}\nu_{23R} - m_{23}\bar{\nu}_{23L}\nu_{23R} - m_{13}\bar{\nu}_{13L}\nu_{13R} - \fr 1 2 M\bar{\nu}_{3R}^c\nu_{3R}\crn
&&+ \mathrm{H.c.} \eea
In the basis $(\nu_{aL}, \nu_{13L}, \nu_{23L}, \nu_{13R}^c, \nu_{23R}^c, \nu_{3R}^c)$ we obtain
\bea M_D &=& -\frac{v}{\sqrt2} \begin{pmatrix}
0 & 0 & h_1 & 0 & 0\\ 0 & 0 & 0 & h_2 & 0\\ 0 & 0 & 0 & 0 & h_3
\end{pmatrix},\\
M_M &=& -\frac{1}{\sqrt2} \begin{pmatrix}
0 & 0 & m_{13} & h_4v_{12} & 0\\ 0 & 0 & h_5v_{12} & m_{23} & h_6v_{23}\\ m_{13} & h_5v_{12} & 0 & 0 & 0 \\ h_4v_{12} & m_{23} & 0 & 0 & h_7v_{23} \\ 0 & h_6v_{23} & 0 & h_7v_{23} & \sqrt2 M
\end{pmatrix}. \eea
It can be verified that the resulting light-neutrino mass matrix, $M_\nu\simeq -M_DM_M^{-1}M_D^T$, has rank one, implying that only a single neutrino acquires mass at tree level. 
This property follows from the structure of $M_D$, where only one effective combination of heavy states couples to the SM neutrinos, thereby limiting the rank of the seesaw-induced mass matrix.

For illustration, we take $m_{13}\simeq v_{12}\simeq\mathcal{O}(10^{5})$~GeV, $m_{23}\simeq v_{23}\simeq\mathcal{O}(10^{4})$~GeV, $M\simeq\mathcal{O}(10^{14})$~GeV, while the Yukawa couplings $h_{1,\cdots,7}$ are assumed to be of $\mathcal{O}(1)$. The resulting light neutrino mass matrix is approximately given by
\be M_\nu\simeq \begin{pmatrix}
0.0100835 & -0.0104394 & -0.045229\\
-0.0104394 & 0.010808 & 0.0468258\\
-0.045229 & 0.0468258 & 0.202873
\end{pmatrix}\text{ eV}. \ee
This matrix reproduces the observed large leptonic mixing angles~\cite{ParticleDataGroup:2024cfk}.

To lift the rank of the neutrino mass matrix, we extend the scalar sector by introducing inert fields that are odd under the residual $\mathbb{Z}_2$ symmetry, in the same way as the right-handed neutrinos $\nu_{1R}$ and $\nu_{2R}$. One possible choice is
\be\eta=\begin{pmatrix}
\eta^0\\\eta^-
\end{pmatrix}\sim(1,2,0,0,-1/2,-1),\hs \rho\sim(1,1,0,0,0,1), \ee
which induce couplings
\bea \mathcal{L} &\supset& \kappa_{31}\bar{l}_{3L}\eta\nu_{1R}+\mu H\eta\rho+\mu'\rho\rho\chi^*+ \mathrm{H.c.},\label{yuneu} \eea 
where $\kappa_{31}$ is a dimensionless coupling, while $\mu$ and $\mu'$ have mass dimension. Alternatively, one may consider two scalar doublets,
 \be\eta=\begin{pmatrix}
\eta^0\\\eta^-
\end{pmatrix}\sim(1,2,0,0,-1/2,-1),\hs \rho=\begin{pmatrix}
\rho^0\\\rho^-
\end{pmatrix}\sim(1,2,0,0,-1/2,1), \ee
which lead to the interactions
\bea \mathcal{L} &\supset& \kappa_{31}\bar{l}_{3L}\eta\nu_{1R}+\kappa_{32}\bar{l}_{3L}\rho\nu_{2R}+\la' HH\eta\rho+\mu'\eta^\dag\rho\chi^*+ \mathrm{H.c.}\label{yuneu} \eea 
These interactions generate one-loop corrections to the neutrino mass matrix. At the schematic level, the loop-induced contribution can be estimated as
\be (M_\nu)_{33}^{\rm loop} \sim \frac{\ka^2M_f}{32\pi^2}\left(\frac{m_R^2}{M_f^2-m_R^2} \ln\frac{M_f^2}{m_R^2}-\frac{m_I^2}{M_f^2-m_I^2} \ln\frac{M_f^2}{m_I^2}\right),
\ee
where $\ka\sim\ka_{31,32}$, and $M_f$, $m_R$, $m_I$ denote the masses of the heavy Majorana fermions and the CP-even and CP-odd scalar components in the loop. In the limit of small scalar mass splitting, this expression reduces to
\be (M_\nu)_{33}^{\rm loop}\sim \frac{\ka^2}{16\pi^2}\,\frac{\Delta m^2}{M_f},
\ee
with $\Delta m^2 \equiv m_R^2 - m_I^2$, induced by scalar mixing after symmetry breaking, e.g. $\Delta m^2 \sim \mu v$ (singlet $\rho$) or $\la' v^2$ (two doublets). This form makes the loop suppression explicit and shows that sub-eV neutrino masses can be obtained for heavy fermions above the electroweak scale. 

Compared with the standard scotogenic scenario, where neutrino masses arise purely at the loop level, the present framework combines a rank-one tree-level seesaw contribution with radiative corrections. The latter lift the rank of the neutrino mass matrix, yielding a rank-two structure. Consequently, the full neutrino mass matrix contains two nonzero neutrino masses, consistent with current oscillation data~\cite{ParticleDataGroup:2024cfk,Kajita:2016cak, McDonald:2016ixn}.

\section{\label{dm}Dark matter}
In the present framework, the residual dark parity guarantees the stability of the lightest particle in the dark sector. The DM candidate may therefore be either a fermion or a scalar. Its relic abundance is determined by interactions mediated by the dark gauge boson and/or by scalar portals connecting the visible and dark sectors.

\subsection{Fermionic dark matter}
We first consider the possibility that the lightest $\mathbb{Z}_2$-odd particle is a fermion, which can be identified with the right-handed neutrino $\nu_{1R}$. In this case, $\nu_{1R}$ becomes a stable DM candidate. Its dominant annihilation channels include
\be \nu_{1R}\nu_{1R}\to ff^c,hh,ZZ,W^+W^-,Z'Z',h'Z',h'h', \ee
where $f$ denotes SM fermions, while $h$, $Z$, and $W^\pm$ are the SM Higgs and electroweak gauge bosons. The particles $Z'$ and $h'$ are new bosons associated with the $U(1)_D$ gauge symmetry. The elastic scattering of $\nu_{1R}$ with nucleons proceeds through Higgs and gauge-boson mediation. The DM phenomenology of this scenario has been studied in detail in Refs.~\cite{VanLoi:2025fmy, VanDong:2023xbd, VanDong:2024lry}. 

Assuming the $h'$ portal dominates the annihilation process, the thermally averaged cross section is set by the $s$-channel scalar exchange and can be schematically written as $\langle\sigma v\rangle \sim M_{\nu_{1R}}^2/(4M_{\nu_{1R}}^2-m_{h'}^2)^2$, up to couplings of order unity. The observed relic abundance $\Omega_{\nu_{1R}} h^2 \simeq 0.12$ is reproduced for the typical freeze-out value $\langle\sigma v\rangle \sim 3\times 10^{-26}\,\text{cm}^3/\text{s}$~\cite{Planck:2018vyg}. For direct detection, the spin-independent (SI) cross section mediated by scalar mixing between $h$ and $h'$ induced by $\varphi$ is estimated to be $\sigma^\text{SI}_{\nu_{1R}\text{--}N} \sim 10^{-46}\,\text{cm}^2$, for a TeV-scale $\nu_{1R}$ and a representative mixing angle of order $10^{-2}$~\cite{XENON:2023cxc, PandaX:2024qfu, LZ:2024zvo}.

\subsection{Scalar dark matter}
Alternatively, the lightest $\mathbb{Z}_2$-odd particle may be a scalar field $X$, which can serve as a viable DM candidate. The dominant annihilation channels include
\be XX\to ff^c,hh,ZZ,W^+W^-,Z'Z',h'h',\ee
while the elastic scattering of $X$ with nucleons proceeds mainly through the SM Higgs portal. As shown in Refs.~\cite{VanDong:2024lry,deBoer:2021pon, VanLoi:2026vqn}, a TeV-scale scalar $X$ can reproduce the observed relic abundance for the typical freeze-out value $\langle\sigma v\rangle \sim 3\times10^{-26}\,\text{cm}^3/\text{s}$~\cite{Planck:2018vyg}, via $s$-channel $h'$-mediated annihilation into SM particles. The SI direct-detection cross section mediated by the Higgs $h$ can reach $\sigma^\text{SI}_{X\text{--}N} \sim 10^{-46}\,\text{cm}^2$, for a TeV-scale $X$ scalar and a $X$--$h$ coupling of order $10^{-2}$.

\section{\label{conc}Conclusion}

In this work, we have studied a framework with generation-separated hypercharges and an additional dark gauge symmetry $U(1)_D$. After the extended hypercharge symmetry is spontaneously broken to the SM $U(1)_Y$ group, hierarchical masses of charged fermions can arise from higher-dimensional operators involving hyperon fields. Focusing on a realization with four hyperons, we have examined neutrino mass generation through either a seesaw or a scotoseesaw mechanism, both of which can accommodate large leptonic mixing. The breaking of $U(1)_D$ leaves a residual $\mathbb{Z}_2$ symmetry that stabilizes DM, allowing either a fermionic or a scalar candidate with viable TeV-scale parameter regions consistent with cosmological observations and current experimental constraints. Overall, this framework provides a simple and unified setting that connects flavor structure, neutrino masses, and DM.

\bibliographystyle{JHEP}

\bibliography{TriYTriD}

\end{document}